\documentclass{ws-procs9x6-cpt19}

\begin{document}

\newcommand{\refeq}[1]{(\ref{#1})}
\def\etal {{\it et al.}}
%any other macros go here 
\def\hydrogen{H}
\def\antihydrogen{$\overline{\mathrm{H}}$}

\title{Rabi Experiments on the $\sigma$ and $\pi$ Hyperfine Transitions in Hydrogen and Status of ASACUSA's Antihydrogen Program}

\author{M.C.\ Simon}

\address{Stefan Meyer Institute, Austrian Academy of Sciences,\\
Boltzmanngasse 3, 1090 Wien, Austria}

\author{
On behalf of the ASACUSA Collaboration\footnote{\url{http://asacusa.web.cern.ch/ASACUSA/asacusaweb/members_files/members.shtml}}
}

\begin{abstract}
We report on the status of the in-beam hyperfine-structure measurements 
on ground-state antihydrogen by ASACUSA 
and on recent results obtained in supporting measurements from hydrogen.
The $\sigma_1$ and $\pi_1$ transitions can now be investigated, 
which is beneficial from both theoretical and experimental perspectives.
We discuss systematic effects from resonance interference 
originating from the chosen field geometries in the interaction region, 
and how their impact can be managed 
by appropriate data-taking or design concepts.
\end{abstract}

\bodymatter

\section{Status of ASACUSA's antihydrogen program}

ASACUSA (Atomic Spectroscopy And Collisions Using Slow Antiprotons) 
is one of several collaborations 
studying antimatter at the antiproton decelerator at CERN.
The majority of experiments in this area compare antimatter properties 
to those of their matter counterparts 
for a precise test of CPT symmetry.
Recent results on the magnetic moment of the antiproton\cite{BASE} 
by BASE
as well as the 1S--2S,\cite{Ahmadi2018-1S2S} 1S--2P,\cite{Ahmadi2018-1S2P}
and hyperfine transitions\cite{Ahmadi2017}
of antihydrogen (\antihydrogen) by ALPHA\cite{ALPHA}
demonstrate the rapid progress in this field.
All existing measurements confirm CPT symmetry 
within their respective uncertainties.
Thus, 
the quest for ever higher precision tests 
by the antiproton-decelerator community continues.
In this spirit, 
ASACUSA is working toward a hyperfine-splitting determination 
based on Rabi spectroscopy.
The SME coefficients, 
which can be tested or constrained by this specific experiment 
have been discussed 
in the proceedings of the previous meeting in this series.\cite{Widmann2016}
The approach followed by ASACUSA 
requires the formation of a beam of \antihydrogen\ 
and offers the advantage that the actual measurement, 
i.e., 
the interaction with microwaves, 
takes place in a well-controlled environment 
far away from the strong fields and field gradients 
of the trap for \antihydrogen\ formation.\cite{Widmann2013,Mohri2003} 
The observation of extracted antiatoms 
had been reported in Ref.\ \refcite{Kuroda2014}, 
and later this beam had its quantum-state distribution 
characterized in Ref.\ \refcite{Malbrunot2018}.
Ground-state \antihydrogen\ has clearly been observed. 
However, 
a rate increase is still needed 
before spectroscopic measurements with reasonable acquisition times can commence.

Due to the current long shutdown at CERN, 
antiproton physics is on hold until 2021.
Then, 
the Extra-Low ENergy Antiproton ring (ELENA) 
will be in operation and the \antihydrogen\  experiment of ASACUSA 
will receive a dedicated beamline.
Therefore, 
the multi-trap setup does not have to be removed from the zone anymore 
in order to make space for other antimatter studies pursued by ASACUSA.
Currently, 
the setup is being installed at its final position 
and matter studies 
(i.e., mixing of protons and electrons) 
are planned to continue optimizing the formation process during the shutdown.

\section{Transitions within the hyperfine sublevels}

The two transitions that occur between a low-field seeking triplet state 
($F \! = \! 1$) 
and the singlet state 
($F \! = \! 0$) 
are called $\sigma_1$ and $\pi_1$ transitions 
(see Table\ \ref{tab:hfsublevels}).
Those are accessible in a Rabi-type experiment 
and approach the value of interest at zero field.
In the interaction volume, 
one has to provide both an oscillating magnetic field $B_\text{osc}$ 
at $1.42\,$GHz to stimulate hyperfine transitions, 
and an external static magnetic field $B_\text{ext}$ 
to control the Zeeman shift.
A key difference between the two transitions is 
that they need different relative orientations of $B_\text{osc}$ and $B_\text{ext}$.
The component of $B_\text{osc}$ aligned parallel to $B_\text{ext}$ 
stimulates the $\sigma_1$ transition, 
and the orthogonal one the $\pi_1$ transition.
In the ASACUSA setup, 
these fields are provided by a cavity of strip-line geometry 
with a large acceptance and surrounding coils.

\begin{table}
\tbl{Properties of the hyperfine sublevels of ground-state \antihydrogen . }
{
\begin{tabular}{@{}rllcc@{}}
\toprule
state: $ | F, M_F \rangle $ & \ Zeeman shift                & Stern--Gerlach  & $\sigma_1$ & $\pi_1$ \\
\colrule
$ | \ 1,-1 \ \rangle $                 & \ 1$^\text{st}O$ \ \ (linear)    &  low-field seeker &         & initial \\
$ | \ 1, \hspace{0.23cm} 0 \ \rangle $ & \ 2$^\text{nd}O$ \ (hyperbolic)  &  low-field seeker & initial &         \\
$ | \ 1, \hspace{0.23cm} 1 \ \rangle $ & \ 1$^\text{st}O$ \ \ (linear)    & high-field seeker &         &         \\
$ | \ 0, \hspace{0.23cm} 0 \ \rangle $ & \ 2$^\text{nd}O$ \ (hyperbolic)  & high-field seeker & final   & final   \\
\botrule
\end{tabular}
}
\label{tab:hfsublevels}
\end{table}

\begin{figure}
\begin{center}
\includegraphics[width=4.3in]{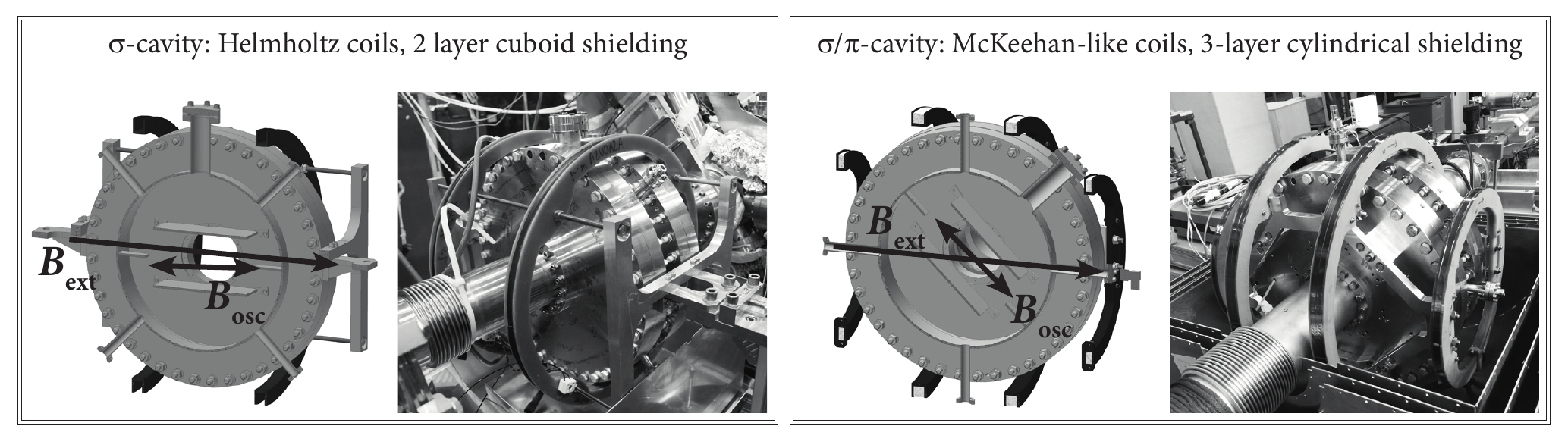}
\end{center}
\caption{ASACUSA's devices for providing an oscillating $B_\text{osc}$ and external static magnetic field $B_\text{ext}$ in the interaction region of the Rabi experiment using microwave cavities and coil configurations.
}
\label{fig:cavities}
\end{figure}

In the low-field regime, 
the $\sigma_1$ transition has a weak magnetic-field sensitivity 
making it more robust against systematic effects.
However, 
it is~also insensitive to SME coefficients,\cite{Kostelecky2015}
which relates to the fact that $\Delta M_F \! = \! 0$.
In contrast, 
the $\pi_1$ transition probes the nonrelativistic spherical SME coeffients
${g_w}_{(2q)10}^{\text{NR}(0\text{B})}$, ${g_w}_{(2q)10}^{\text{NR}(1\text{B})}$ 
(CPT odd) 
and ${H_w}_{(2q)10}^{\text{NR}(0\text{B})}$, ${H_w}_{(2q)10}^{\text{NR}(1\text{B})}$ (CPT even).\cite{Widmann2016,Kostelecky2015}
Control over systematic effects 
is a much stronger experimental challenge 
due to the first-order shift with $B_\text{ext}$ of $14\,$Hz/nT.

Helmholtz coils produce a sufficiently homogeneous static field 
to investigate the $\sigma_1$ transition 
($\sigma_{|B|} / \bar{|B|} < 5 \%$).
The cavity providing the oscillating field 
is placed within those coils in such a way 
that the required parallel alignment is achieved, 
as shown in the left panel of Fig.\ \ref{fig:cavities}.
For more details and a description of the entire \hydrogen -beam setup 
for commissioning the hyperfine spectrometer, 
see Ref.\ \refcite{Malbrunot2019}.
A determination of the \hydrogen\ hyperfine structure with $2.7\,$ppm precision 
in agreement with the literature value 
was accomplished\cite{Diermaier2017} by measuring the $\sigma_1$ transition 
at various static-field values 
and extrapolating to zero field. 
However, 
as mentioned before the $\sigma_1$ transition 
is not the first choice for a CPT test, 
which motivated upgrades of the interaction apparatus 
for investigations of the $\pi_1$ transition.

We designed McKeehan-like coils\cite{McKeehan1936} 
to provide a more homogeneous magnetic field 
($\sigma_{|B|} / \bar{|B|} < 0.1 \%$).
The cavity can be rotated in steps of $45^\circ$ 
within the coil arrangement 
thereby providing flexible alignment of $B_\text{osc}$ to $B_\text{ext}$.
A $45^\circ$ alignment enables simultaneous access to both transitions.
The upgraded device is shown in the right panel of Fig.~\ref{fig:cavities}.

\section{Interference effect and future measurements and devices}

At small $B_\text{ext}$, 
the separation of the $\sigma_1$ and $\pi_1$ transitions 
becomes comparable to the line widths, 
which is on the order of $10\,$kHz 
as the interaction time is typically $100\,\mu$s 
(beam velocity $\sim 1\,$km/s, 
cavity length $\sim 100\,$mm).
For example, 
at $4.6\,\mu$T 
the separation decreases to $65\,$kHz, 
and the interference leads to asymmetric line shapes 
and systematic shifts of the extracted central frequencies, 
if one applies the symmetric fit function 
for a two-level system in this regime.
Currently, 
we apply corrections for the effect, 
and the development of a fit procedure 
based on the complete four-level system of the ground-state hyperfine states 
is under consideration.
On the other hand, 
the interference effect and resulting systematic shifts can be avoided 
with purely parallel or orthogonal alignment of the fields.
This clean solution will be realized in a Ramsey apparatus 
that is currently in its final design phase.\cite{HyDRA}
In principle, 
the respective alignment can also be adjusted in steps of $45^\circ$ 
with the present Rabi apparatus. 
However, 
a change requires breaking the vacuum.
In this design, 
the advantageous opportunity 
for interleaved measurements of $\pi_1$ and $\sigma_1$ transitions 
is tied to the alignment of the fields 
that also gives rise to the interference effect.
By operating at sufficiently high $B_\text{ext}$ 
(e.g., $23/69/115\,\mu$T as in Ref.\ \refcite{Widmann2018}) 
the systematic shifts and related uncertainties 
can easily be kept below $\sim 10\,$ppb, 
an acceptable level 
for the anticipated first-stage results on \antihydrogen .
A measurement of a single pair of a $\pi_1$ and $\sigma_1$ transitions 
in this regime of $B_\text{ext}$ 
provides a reliable way to determine the zero-field splitting 
from a calculation,\cite{Widmann2018} 
i.e., 
without extrapolating to zero field by a fit.
Thus, 
the present apparatus is well suited 
for a first Rabi-type SME-sensitive \antihydrogen\ hyperfine-structure measurement.

\section*{Acknowledgments}
% We thank Dr. Fritz Caspers for excellent support.
This work was funded by the European Research Council (grant no.\ 291242), 
the Austrian Ministry of Science and Research, and the Austrian Science Fund (FWF) through DKPI (W1252).

\end{document}